\documentstyle[sprocl]{article}

\input{psfig}
\def\be{\begin{equation}}
\def\ee{\end{equation}}
\def\bea{\begin{eqnarray}}
\def\eea{\end{eqnarray}}

\begin{document}

\title{Extracting the equation of state
	from a microscopic non-equilibrium model}

\author{C.~Spieles\footnote[1]{Participant in the PANIC'96 Conference}, 
A.~Dumitru$^*$,
S.~A.~Bass, M.~Bleicher, J.~Brachmann, M.~Brandstetter,
C.~Ernst, L.~Gerland, 
J.~Konopka, S.~Soff, H.~Weber, L.~A.~Winckelmann, J.~Maruhn, 
H.~St\"ocker, W.~Greiner}

\address{Institut f\"ur Theoretische Physik\\
Johann Wolfgang Goethe Universit\"at\\
    60054 Frankfurt am Main, Germany}

\maketitle

One of the main goals of relativistic heavy ion collisions 
is the determination of the nuclear equation of state. 
At high energies,
semiclassical cascade models in terms of scattering hadrons
 have proven to be rather
accurate in explaining experimental data. Therefore it is of 
fundamental interest to extract
\begin{figure}[h]
\vspace*{-2cm}
\begin{minipage}[b]{2.3in}
\vspace*{-13cm}
\hspace*{0.2cm}
\centerline{\psfig{figure=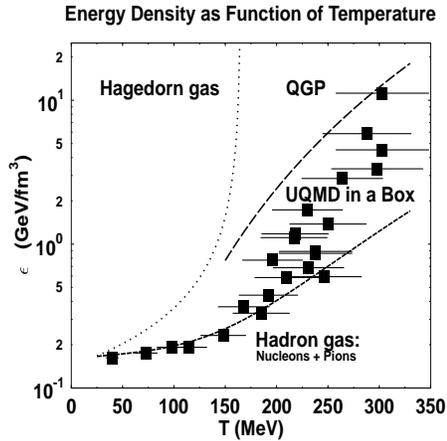,width=6cm,height=6cm}}
\caption{\label{fig:eos}
'Equation of state' of infinite nuclear matter, calculated with URQMD
(preliminary). Shown is the energy density as a function of temperature
(extracted with a least square fit to the baryon momentum spectrum) at 
fixed net-baryon
density of $\rho=0.16\; {\rm fm^{-3}}$. The equations of state of a
Hagedorn-gas, of a quark-gluon plasma, and of an ideal gas of nucleons and
ultrarelativistic pions are also depicted.}
\end{minipage}
\hfill
\begin{minipage}[b]{2.2in}
\vspace*{1.4cm}
the equation of state from such a microscopic model, i.~e. to
investigate the equilibrium limits and bulk properties, which are not
an explicit input to the non-equilibrium transport approach with its
complicated collision term (unlike e.~g. in hydrodynamics).
In Fig.~\ref{fig:eos} we study the thermodynamic properties of
infinite nuclear matter with the Ultrarelativistic Quantum Molecular 
Dynamics 1.0$\beta$ (URQMD), 
a semiclassical transport model\cite{uqmd}.
The model is based on
classical propagation of hadrons and stochastic scattering ($s$ channel
excitation of baryonic and mesonic resonances/strings, $t$ channel
excitation, deexcitation and decay). For this study the potentials have been
switched off. To simulate infinite hadronic systems we construct
 a box of volume 250~fm$^3$ with periodic boundary conditions, and 
initialize 40 nucleons --- that is
\end{minipage}
\end{figure}
\clearpage
\noindent
ground-state nuclear density --- randomly in phase
space, varying the total energy density. 
After the system has equilibrated according to the
URQMD simulation, the temperature is extracted by fitting the particles'
momentum spectra. Alternatively, the temperature can be extracted from the
relative abundances of different hadrons, e.~g. the $\Delta/N$ ratio, which
should yield the same temperature. This has been checked in the current
simulation.
The result of this procedure is plotted in
Fig.~\ref{fig:eos}. It appears that the energy density rises faster than
$T^4$ at high temperatures of $T\approx 200-300$~MeV. This indicates an
increase in the number of degrees of freedom. It may be interpreted as a
consequence of the numerous high mass resonances and string excitations,
which in a way  release constituent quark degrees of freedom (but, of
course, no free current quarks as in an ideal QGP). Investigations
of equilibration times and relative particle and cluster abundances are in
progress. Moreover, the admittedly poor statistics have to be improved.

\begin{figure}[t]
\begin{minipage}[b]{2in}
\vskip 0mm
\vspace{0.7cm}
\hspace*{-0.25cm}
\psfig{figure=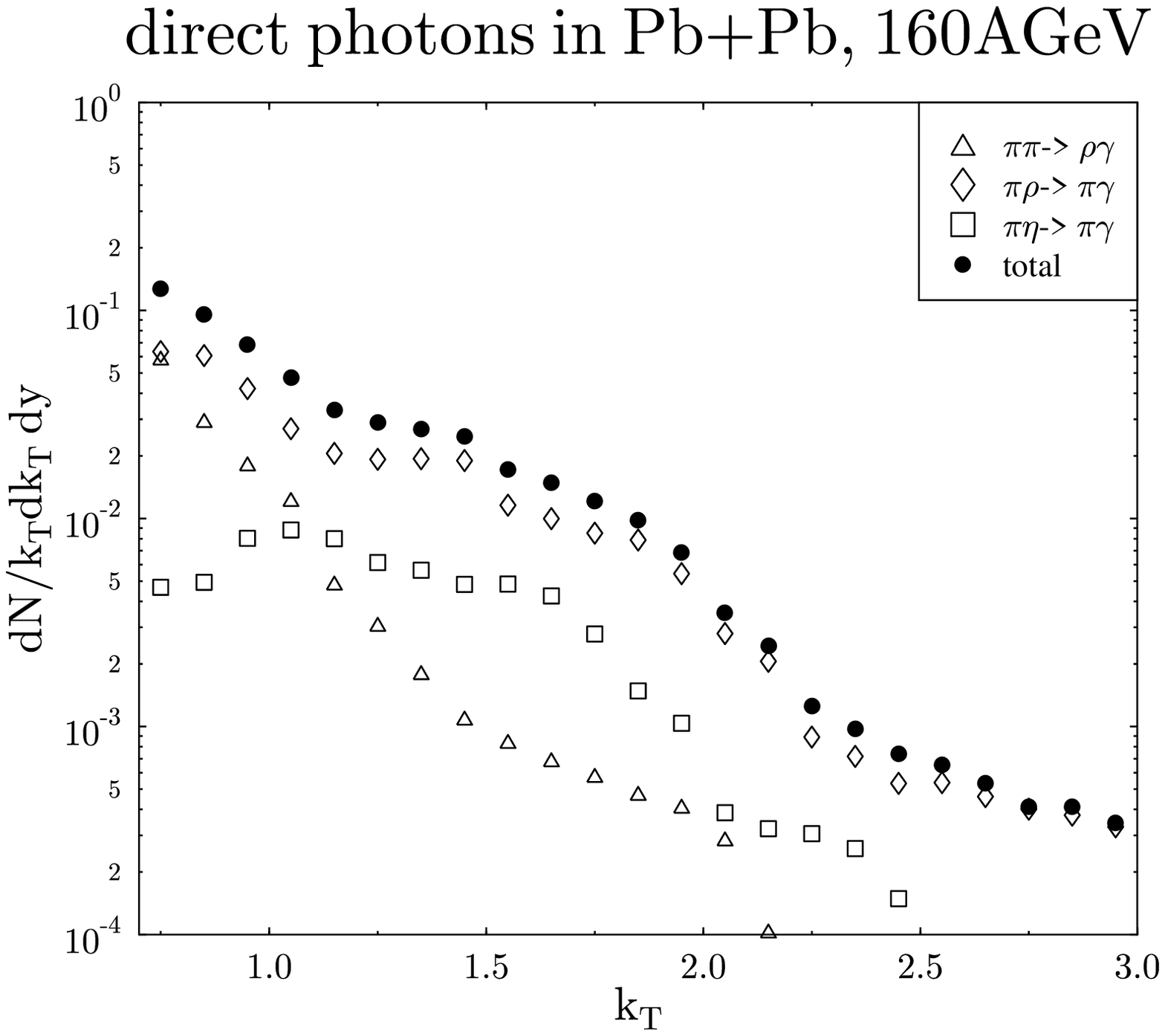,width=2.5in}
\vskip 2mm
\vspace{0.0cm}
\end{minipage}
\hfill
\begin{minipage}[b]{2in}
\vskip 0mm
\vspace{0.cm}
\hspace*{-1.cm}
\psfig{figure=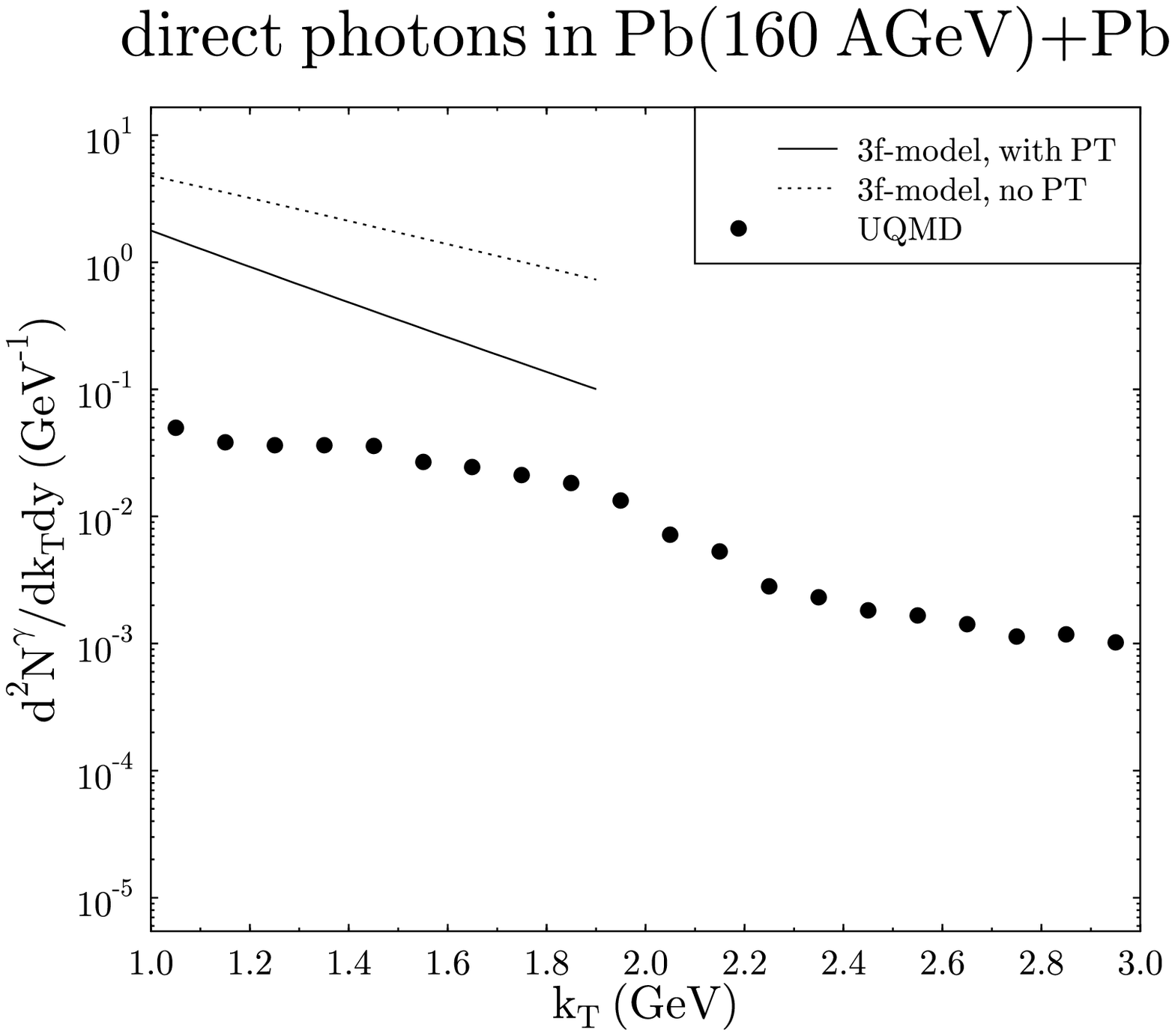,width=2.5in}
\vskip 2mm
\vspace{0.0cm}
\end{minipage}
\caption{Transverse momentum spectrum of directly produced photons in 
Pb+Pb collisions at 160~GeV/u calculated with URQMD (left). The
contributions of the different processes are shown. The resulting spectrum
is compared with hydrodynamical calculations (right). For the photon source,
an equation of state with and without a phase transition is assumed.
\label{photon}}
\vspace*{-0.5cm}
\end{figure}

Experimentally, one can access the equation of state of strongly interacting
matter by measuring thermal electromagnetic radiation emitted in heavy ion
collisions.
We have calculated direct photon production in Pb+Pb collisions at
160~GeV/u within the framework of URQMD~1.0~$\beta$. To improve statistics,
the total meson-meson cross sections were fixed to 15~mb (independent of
$\sqrt{s}$). For the sake of comparison with earlier hydrodynamical
calculations (see \cite{ADRIAN} and refs.~therein) we considered only the processes 
$\pi\rho\to\pi\gamma$, $\pi\eta\to\pi\gamma$ 
and $\pi\pi\to\rho\gamma$. The amplitudes are taken from \cite{KLS}. 
As can be seen in Fig.~\ref{photon}~(left) the 
$\pi+\rho\to\pi+\gamma$ process dominates the
spectrum in the transverse momentum range considered.
Fig.~\ref{photon}~(right) shows that the
direct photon slope from the microscopic calculation equals that from the
hydrodynamical calculation without a phase transition in the equation of
state of the photon source. In the case of a phase transition the slope becomes significantly steeper due to a lower temperature at
given energy density. The lower absolute yields of the direct
photons from the URQMD may indicate that the meson abundance at early times
is smaller than in the hydrodynamical model of \cite{ADRIAN}, although 
the final
yields are similar  (due to higher
mass mesons and baryon resonances). Thermal
and in particular chemical equilibrium may not be established within URQMD
in the early stage of the collision (where the large transverse 
momentum photons are produced). In contrast, in the hydrodynamical calculation
local thermal and chemical equilibrium of the produced particles is assumed
from the very beginning.
Experimental studies on direct photons, under way at
CERN~\cite{wa98}, will help to settle these questions in the near future.

\section*{Acknowledgments}
This work was supported by the
GSI, Darmstadt, the BMBF, Bonn, Deutsche Forschungsgemeinschaft,
and the Graduiertenkolleg Schwer\-ionen\-physik (Frankfurt/Giessen).


\section*{References}
\begin{thebibliography}{99}
\bibitem{uqmd} S.~A.~Bass, M.~Bleicher, M.~Brandstetter, 
C.~Ernst, L.~Gerland, C.~Hartnack,
J.~Konopka, S.~Soff, C.~Spieles, H.~Weber, L.~A.~Winckelmann,
J. Aichelin, N.~Amelin, H.~St\"ocker and W.~Greiner:
source code and technical documentation, to be published;\\
L.A.~Winckelmann et al., Proc. of the 12th Int. Conf. on Ultrarelativistic
Nucleus-Nucleus Collisions, Quark Matter '96, Heidelberg, May 20-24 (1996)
\bibitem{ADRIAN}A. Dumitru, U. Katscher, J.A. Maruhn, H. St\"ocker,
W. Greiner, D.H. Rischke: Phys. Rev. C51 (1995) 2166
\bibitem{KLS} J. Kapusta, P. Lichard, D. Seibert: Phys. Rev. D44 (1991) 2774
\bibitem{wa98} T.~Peitzmann for the WA98 collab.: Talk given at the 
12th Int. Conf. on Ultrarelativistic
Nucleus-Nucleus Collisions, Quark Matter '96, Heidelberg, May 20-24 (1996)

\end{thebibliography}
\end{document}